\documentclass[preprint,12pt,5p]{elsarticle}

\usepackage[italian,english]{babel}
\usepackage{hyperref}
\usepackage{ifpdf}
\usepackage{subfigure}

\usepackage{amssymb}
\usepackage{amsfonts}
\usepackage{epsf}
\usepackage{rotating}
\usepackage{graphicx}
\usepackage{amsmath}
\usepackage{fancyhdr}
\usepackage{lineno}

\usepackage{babel}
\usepackage{graphics}
\usepackage{pstricks}
\usepackage{color}
\usepackage{multirow}

\def\beq{\begin{equation}}
\def\eeq{\end{equation}}
\def\bea{\begin{eqnarray}}
\def\eea{\end{eqnarray}}
\def\nn{\nonumber}
\def\bei{\begin{itemize}}
\def\eei{\end{itemize}}
\def\bmat{\begin{matrix}}
\def\emat{\end{matrix}}
\def\={\,=\,}
\def\+{\,+\,}
\def\-{\,-\,}
\def\ca{c_\alpha}
\def\sa{s_\alpha}
\def\cb{c_\beta}
\def\sb{s_\beta}

\def\ptmiss{\not\!\!{p_T}}

\begin{document}


\title{Scalar Dark Matter in Leptophilic Two-Higgs-Doublet Model}

\author[a]{Priyotosh Bandyopadhyay}
\ead{bpriyo@iith.ac.in} 

\author[b]{Eung Jin Chun}
\ead{ejchun@kias.re.kr} 

\author[c]{Rusa Mandal}
\ead{rusam@imsc.res.in}

\address[a]{Indian Institute of Technology Hyderabad, Kandi,  Sangareddy-502287, Telengana, India}
\address[b]{Korea Institute for Advanced Study, Seoul 130-722, Korea}

\address[c]{The 
	Institute  of Mathematical Sciences, HBNI, Taramani, Chennai 600113, India}

	

\begin{abstract}
Two-Higgs-Doublet Model of Type-X in the large $\tan\beta$ limit becomes leptophilic to allow a light 
pseudo-scalar $A$ and thus provides an explanation of the muon $g-2$ anomaly.  
Introducing a singlet scalar dark matter $S$ in this context, one finds that 
two important dark matter properties, nucleonic scattering and self-annihilation, are featured separately  by individual couplings of dark matter to the two Higgs doublets.   
While one of the two couplings is strongly constrained by direct detection experiments, 
the other remains free to be adjusted for the relic density mainly through the process $SS\to AA$. This leads to the $4\tau$ final states  which can be probed by galactic gamma ray detections. 
	
\end{abstract}

\begin{keyword}
	Two-Higgs-Doublet Model \sep Dark Matter \sep Direct and Indirect Detection
	
	
	
\end{keyword}


\maketitle
\flushbottom

\section{Introduction}

The existence of dark matter (DM) is supported by various astrophysical and cosmological observations in different gravitational length scales. The best candidate for dark matter is a stable neutral particle beyond the Standard Model (SM).  The simplest working model is to extend the SM by adding a singlet scalar \cite{Silveira:1985rk,McDonald:1993ex} and thus allowing its coupling to the SM Higgs doublet which determines the microscopic properties of the dark matter particle. This idea of Higgs portal  has been very popular in recent years and studied extensively by many authors \cite{Athron:2017kgt}. However, such a simplistic scenario is tightly constrained by the current direct detection experiments since a single Higgs portal coupling determines 
both the thermal relic density and the DM-nucleon scattering rate.

One is then tempted to study the scalar dark matter property in  popular Two-Higgs-Doublet Models (2HDMs)
\cite{Gunion:2002zf}. Having more degrees of freedom, 
two independent Higgs portal couplings and extra Higgs bosons,
one could find a large parameter space accommodating the current experimental limits and 
enriching phenomenological consequences \cite{2hdmDM}.

\medskip

The purpose of this work is to realize a scalar singlet DM through Higgs portal in the context of a specific 2HDM which can accommodate the observed deviation of the muon $g-2$.  Among four types of $Z_2$-symmetric 2HDMs, the type-X model is found to be a unique option for the explanation of the muon $g-2$ anomaly \cite{Broggio:2014mna} and the relevant parameter space has been explored more precisely  \cite{Cao:2009as,Wang:2014sda,Abe:2015oca,Chun:2016hzs}. Combined with the lepton universality conditions, one can find a large parameter space allowed at $2\sigma$ favoring $\tan\beta\gtrsim 30$ and  $m_A \ll m_{H,H^\pm} \approx 200-400$ GeV \cite{Chun:2016hzs}.  The model can be tested at the LHC by searching for 
a light pseudo-scalar $A$ through $4\tau$ or $2\mu\, 2\tau$  final states \cite{Chun:2015hsa,Goncalves:2016qhh, Chun:2017yob}.

In the large $\tan\beta$ regime, the SM-like Higgs boson reside mostly on the Higgs doublet with a large VEV.
Therefore its coupling to DM is severely constrained by the direct detection experiments. On the other hand, the other Higgs doublet with a small VEV contains mostly the extra Higgs bosons, the light pseudo-scalar $A$, heavy neutral and charged bosons $H$ and $H^\pm$, and thus its coupling to DM controls the thermal relic density preferably through the annihilation channel $SS\to AA$. 

In Sec.~\ref{sec:model}, we describe the basic structure of the model. In Sec.~\ref{sec:direct} and \ref{sec:relic}, we discuss the consequences of DM-nucleon scattering and DM annihilation which determines the relic density as well as the indirect detection, respectively. We conclude in Sec.~\ref{sec:conclusion}.

\section{L2HDM with a scalar singlet}
\label{sec:model}

\begin{figure*}[!ht]
	\begin{center}
		\mbox{\hskip -20 pt\subfigure[]{\includegraphics[width=0.44\linewidth]{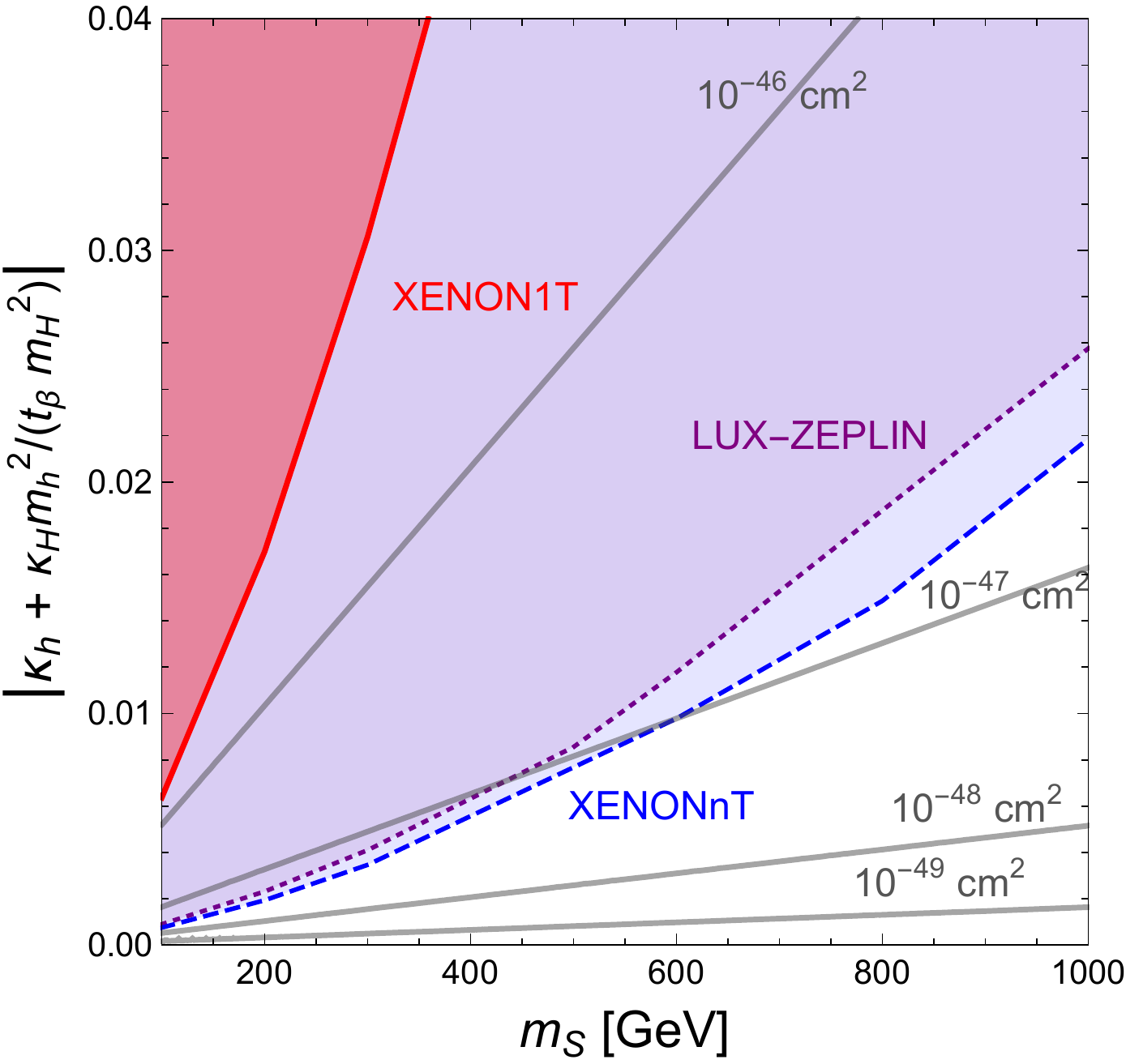}}
			\subfigure[]{\hskip 20 pt \includegraphics[width=0.44\linewidth]{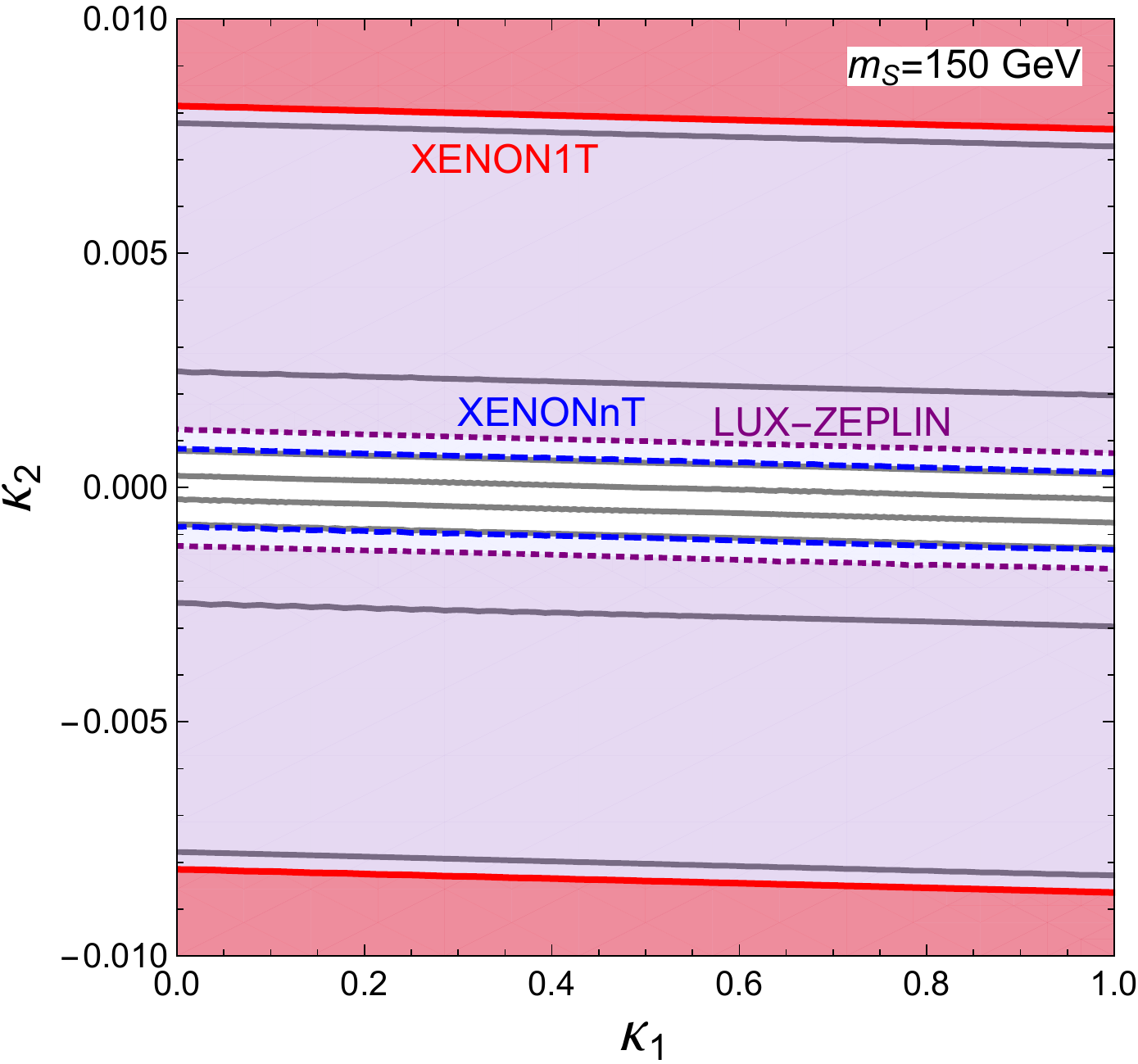}}}
		\caption{(a) The allowed parameter space in the DM mass $m_S$ and the combination of couplings plane for SI scattering cross section. 
			The red solid curve is the current bound from XENON1T~\cite{Aprile:2017iyp} experiment and the purple dot and blue dashed curves are the expected bounds in LUX-ZEPLIN~\cite{Mount:2017qzi} and XENONnT~\cite{Aprile:2015uzo} experiments, respectively. The region above the mentioned curves are excluded at 90\% confidence level.
			(b) The allowed region in $\kappa_1-\kappa_2$ plane is illustrated by choosing $m_S=150$\,GeV from the left panel figure. The color code is the same as of the left panel. We take $m_H=250\,$GeV and $t_\beta=50$ for these plots.
		}
		\label{fig:directD}
	\end{center}
\end{figure*}

Introducing two Higgs doublets $\Phi_{1,2}$ and one singlet scalar $S$ stabilized by the symmetry $S\to -S$,  one can write down the following gauge invariant scalar potential:
\begin{align}  \label{scalar-potential}
V &= m_{11}^2 |\Phi_1|^2
	+ m_{22}^2 |\Phi_2|^2 - m_{12}^2 (\Phi_1^\dagger \Phi_2 + \Phi_1 \Phi_2^\dagger)  \nn  \\
&+ {\lambda_1\over2} |\Phi_1|^4 \!+\! {\lambda_2 \over 2} |\Phi_2|^4 
	\!+\! \lambda_3 |\Phi_1|^2 |\Phi_2|^2 \!+\! \lambda_4 |\Phi_1^\dagger \Phi_2|^2 \nn \\
&	+ {\lambda_5 \over 2}
	\left[ (\Phi_1^\dagger \Phi_2)^2 + (\Phi_1 \Phi_2^\dagger)^2\right]  \nn \\
& +  \frac{1}{2} m_0^2 S^2\! +\! \frac{\lambda_S}{4} S^4 \!+
  \!S^2 \left[ \kappa_1 |\Phi_1|^2 + \kappa_2 |\Phi_2|^2 \right]\!,
\end{align}
where a softly-broken $Z_2$ symmetry is imposed in the 2HDM sector to forbid dangerous flavor violation. 
The model contains four more parameters compared to the usual 2HDMs: one mass parameter $m_0$ and three dimensionless parameters $\lambda_S$ and $\kappa_{1,2}$ for the DM self-coupling and the DM-Higgs couplings, respectively.
Extending the analysis in \cite{Gunion:2002zf}, one can find the following 
simple relations for the vacuum stability~\cite{Klimenko:1984qx}:
\bea
&& \lambda_S>0,\quad  \tilde\lambda_{1} >0,\quad   \tilde\lambda_{2} >0,  \nn\\
&& \tilde\lambda_3>-\sqrt{\tilde\lambda_1 \tilde\lambda_2},  \\ 
&& \tilde\lambda_3 + \lambda_4 - |\lambda_5| > -\sqrt{\tilde\lambda_1 \tilde\lambda_2} \nn 
\eea
where $\tilde\lambda_1\equiv \lambda_1 - \kappa_1^2/2\lambda_S$,  
$\tilde\lambda_2 \equiv \lambda_2 - \kappa_2^2/2\lambda_3$, and 
 $\tilde\lambda_3\equiv \lambda_3 - \kappa_1 \kappa_2/2\lambda_S$.
As we will see, the desired dark matter properties require $|\kappa_{1,2}|\ll 1$ and thus the vacuum stability condition can be easily satisfied in a large parameter space. 

Minimization conditions determine the vacuum expectation values $\langle \Phi^0_{1,2} \rangle \equiv v_{1,2}/\sqrt{2}$ around which the Higgs doublets are expressed as
\begin{equation}
\Phi_{1,2} = \left[\eta^+_{1,2}, {1\over\sqrt{2}}\left(v_{1,2} + \rho_{1,2} + i \eta^0_{1,2}\right)\right].
\end{equation}
Removing the Goldstone modes, there appear five massive fields  denoted by $H^\pm, A, H$ and $h$.
Assuming negligible CP violation, $H^\pm$ and $A$ are given by
\begin{equation}
 H^\pm, A =- s_\beta\, \eta_1^{\pm, 0}  + c_\beta\, \eta_2^{\pm, 0},
\end{equation}
where the angle $\beta$ is determined from $t_\beta\equiv \tan\beta =v_2/v_1$.
The neutral CP-even Higgs bosons are diagonalized by the angle $\alpha$:
\bea
&& h = -\sa \rho_1 +\ca \rho_2, \nn \\
&& H = +\ca \rho_1 + \sa \rho_2 , 
\eea
where $h$ denotes the lighter (125 GeV) state.

Normalizing the Yukawa couplings of the neutral bosons to a fermion $f$  by $m_f/v$ where $v=\sqrt{v_1^2+v_2^2} = 246$ GeV, we have the following Yukawa couplings of the Higgs bosons:
\begin{align}
 -{\cal L}_{Y} &=
\sum_{f=u,d,\ell} {m_f\over v}
\left( y_f^h h \bar{f} f + y_f^H H \bar{f} f - i y_f^A A \bar{f} \gamma_5 f \right) \nn \\
&+ \big[ \sqrt{2} V_{ud} H^+  \bar{u} \left( {m_u\over v} y^A_u P_L  + {m_d \over v} y_d^A P_R\right) d \nn \\
&+\sqrt{2} {m_l \over v} y_\ell^A H^+ \bar{\nu} P_R \ell + {\rm h.c. }\big].
\end{align}
Recall that the type-X 2HDM assigns the $Z_2$ symmetry under which $\Phi_1$ and right-handed leptons are odd; and the other particles are even, and thus $\Phi_2$ couples to all the quarks and $\Phi_1$ to leptons.

As a consequence, one has the normalized Yukawa couplings $y^{h,H,A}_f$ given by
\beq
 \label{yukawas}
\begin{tabular}{cc|cc|cc}
$y_{u,d}^A$ & $y_\ell^A$ & $y_{u,d}^H$ &  $y_\ell^H$ & $y_{u,d}^h$ & $y_\ell^h$\cr
\hline  \rule{0pt}{4ex}
$\pm \displaystyle\frac{1}{t_\beta}$ & $t_\beta$ & $\displaystyle\frac{\sa}{\sb}$ & $\displaystyle\frac{\ca}{\cb}$ & 
 $\displaystyle\frac{\ca}{\sb}$ & $-\displaystyle\frac{\sa}{\cb}$~ \cr
\end{tabular}
\eeq
As the 125 GeV Higgs ($h$) behaves like the SM Higgs boson, we can safely take the alignment limit of $\cos(\beta-\alpha)\approx 0$ and  $|y^h_f| \approx 1$ and  $y^{A,H}_{u,d}\propto 1/t_\beta$ and $y^{A,H}_{l}\propto t_\beta$. Notice that $A$ and $H$ couple dominantly to the tau  in the large $\tan\beta$ limit.

The singlet and doublet scalar couplings are given by
\begin{align}
V= \frac{1}{2} S^2& \big[ 2v (\kappa_h h + \kappa_H H) +\kappa_{hh} h^2 +2 \kappa_{hH} hH  \nn \\
& +\kappa_{HH} H^2
+ \kappa_{AA} (A^2 + 2 H^+ H^-) \big], \nn \\
\mbox{where}~~
\kappa_{h} &= -\kappa_1 \sa\cb +\kappa_2 \ca\sb \approx \kappa_1 \cb^2 +\kappa_2 \sb^2, \nn \\
\kappa_{H} &= +\kappa_1 \ca\cb +\kappa_2 \sa\sb \approx (\kappa_1-\kappa_2) \cb\sb. \nn\\
\kappa_{hh}&=\kappa_1 \sa^2 +\kappa_2 \ca^2 \approx \kappa_1 \cb^2 +\kappa_2 \sb^2 ,\nn \\
\kappa_{hH}&= -(\kappa_1-\kappa_2) \ca\sa\approx (\kappa_1-\kappa_2) \cb\sb ,\nn\\
\kappa_{HH}&=\kappa_1 \ca^2 +\kappa_2 \sa^2 \approx \kappa_1 \sb^2 +\kappa_2 \cb^2 , \nn \\
\kappa_{AA}&=\kappa_1 \sb^2 +\kappa_2 \cb^2, 
\label{kappas}
\end{align}
which shows interesting relations  in the alignment limit: $\kappa_h\approx \kappa_{hh}$, $\kappa_{H}\approx \kappa_{hH}$, and $\kappa_{HH} \approx  \kappa_{AA}$. Furthermore, 
one finds further simplification: $\kappa_{h, hh}\sim \kappa_2$, $\kappa_{H,hH} \sim 0$,
and $\kappa_{AA,HH} \sim \kappa_1$
neglecting small contributions suppressed by $1/t_\beta$. 
This behavior determines the major characteristic of the model.

Before starting our main discussions, let us make a few comments on the LHC probe of the model.
As shown in Eq.~\eqref{yukawas}, the extra Higgs couplings to quarks are proportional to  $1/t_\beta$ and thus their single production is suppressed by $1/t^2_\beta$ compared to the SM Higgs production. For this reason a light $A$ (and $H$) is still allowed
by the direct search of di-tau final state at ATLAS \cite{atlasditau} in the large $\tan\beta$ limit, which also explains the muon $g-2$ anomaly. One can also look for usual electroweak productions of $pp\to HA, H^\pm A$, 
ending up with muti-tau signals  \cite{Chun:2015hsa}, or the SM Higgs production and its exotic decay $h\to AA$ \cite{Chun:2017yob}.  The $pp\to HA$ process is of particular interest in the model under consideration as it could lead to a promising signature of di-tau associated with large missing energy.
Having $\kappa_H \propto 1/t_\beta$, however, the $H\to SS$ process (when allowed kinematically) is highly suppressed in the large $\tan\beta$ limit and thus hardly be probed at the LHC.
The recent bounds on the multi-tau events searched by ATLAS in the case of  the chargino/neutralino production \cite{2taumispt} could be relevant for our model parameter space.
Applying the same cuts, e.g., $\ptmiss>150$ GeV and $p_T^{\tau_1,\tau_2}>50,40$ GeV, to our processes, we find that no events survive for the final states searched in Ref.~\cite{2taumispt}.  This is basically due to the following differences: (i)  the $H^\pm A$ and $HA$ production cross-sections are smaller than the chargino/neutralino production by almost one order of magnitude;  (ii) our processes do not generate large missing energy, and $\tau$'s coming from a light $A$ become too soft to pass the above hard cuts as indicated in Ref.~\cite{Chun:2015hsa}.
We have also checked the recent bounds on $2\ell/3\ell+\ptmiss$ final states with kinematic demands: $p^{\ell}_T \ge 20, 30$ GeV and $\ptmiss \ge 130, 150$ GeV, etc \cite{atlas-conf}. However, in the given parameter space  we have the following branching fraction $\mathcal{B}(H\to A Z) \sim 68\%$, $\mathcal{B}(H \to \tau \tau)\sim 32\%$ and $\mathcal{B}(A \to \tau \tau)\sim 99\%$. The charged Higgs also dominantly decays to $A W^\pm$ ($\sim 70\%$), which makes all the dominant production modes, i.e. $HA$, $HH^\pm$ and $H^\pm A$, insensitive to the search of multi-lepton plus large missing energy final states. Thus the recent bounds on the multi-lepton plus missing energy events motivated to probe supersymmetric signals \cite{atlas-conf} can easily be evaded.

\section{DM-nucleon scattering}
\label{sec:direct}

The spin-independent (SI) nucleonic cross section of the DM is given by 
\beq
\sigma_N = \frac{m_N^2 v^2}{\pi (m_S+m_N)^2}
\!\left( \frac{\kappa_h g_{NNh}}{m_h^2}+  \frac{\kappa_H g_{NNH}}{m_H^2} \right)^2\!\!\!\!,
\eeq
where $g_{NNh}\approx 0.0011$~\cite{Alarcon:2011zs} and $g_{NNH}\approx g_{NNh}/t_\beta$.

In Fig.~\ref{fig:directD}(a), by considering the latest XENON1T bound \cite{Aprile:2017iyp} (red solid) and the future sensitivity of the two experiments LUX-ZEPLIN~\cite{Mount:2017qzi} (purple dotted) and XENONnT~\cite{Aprile:2015uzo} (blue dashed), we highlight the allowed region in the plane of DM mass $m_S$ and the combination of couplings $\bigg| \kappa_h + \displaystyle\frac{\kappa_H}{t_\beta}\frac{m_h^2}{m_H^2} \bigg|$. The shaded region above the mentioned direct detection experiment bounds are excluded at $90\%$ confidence level. 
For further illustration, in Fig.~\ref{fig:directD}(b), we choose a benchmark point $m_S=150\,$GeV and show the allowed parameter space in $\kappa_1-\kappa_2$ plane for $m_H=250\,$GeV and $t_\beta=50$. The color code is the same as in Fig.~\ref{fig:directD}(a). 
Note that in the limit of $t_\beta\gg1$ and $m_H>m_h$, the combined coupling is dominated simply by
$\kappa_2$ and thus strongly constrained as in the SM Higgs portal scenario. 
One can also see that it is not possible to make the combined coupling small through cancellation between two  large couplings.  The other coupling $\kappa_1$ is rather unconstrained and thus this freedom allows us to reproduce the right relic density of dark matter.

\section{DM annihilation}
\label{sec:relic}

\begin{figure}[!ht]
	\begin{center}
		\includegraphics[width=0.9\linewidth]{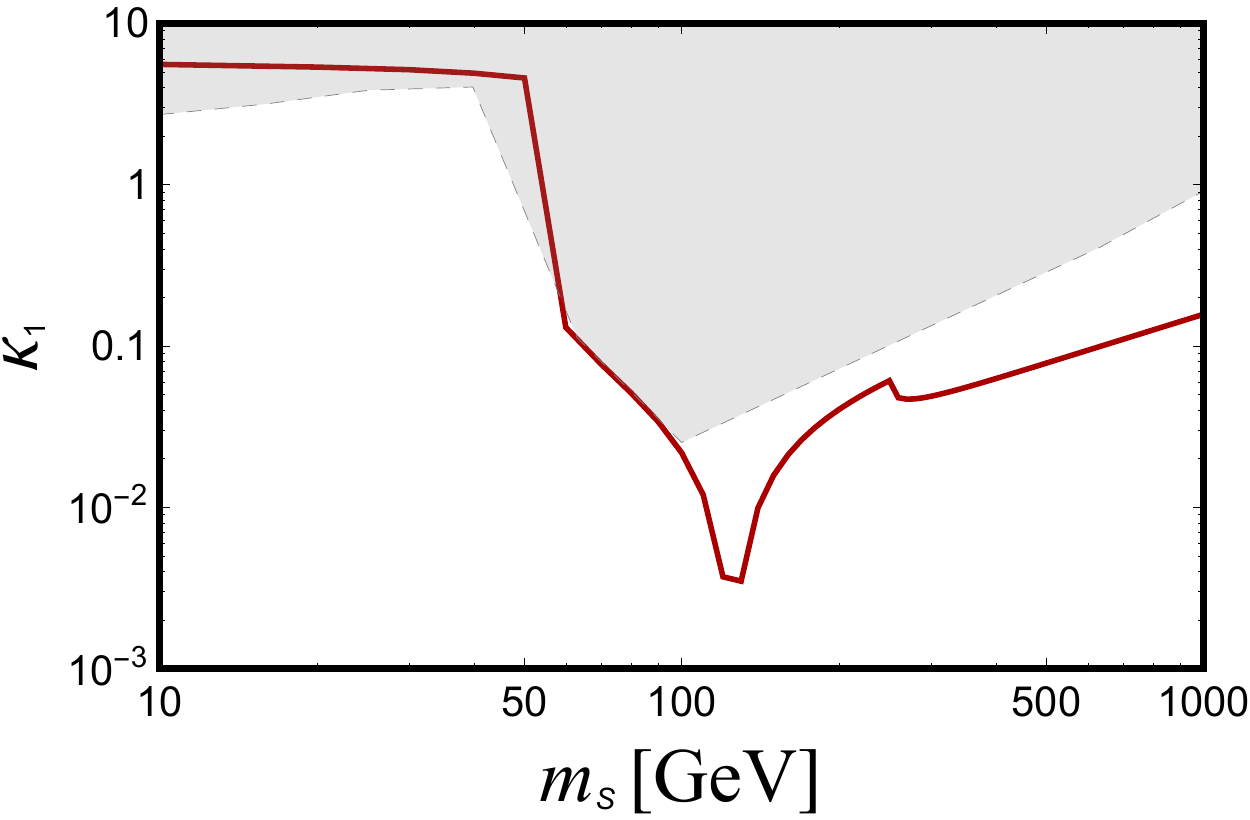}
		\caption{The right DM relic density is obtained by the red line through the DM annihilation channels $SS \to \tau\tau, AA$, and $HH/H^+ H^-$. The gray shaded region is excluded by  Fermi-LAT gamma ray detection in the $2\tau$ \cite{Ackermann:2015zua} and $4\tau$ \cite{Elor:2015bho} final states. The plot is obtained for $m_A=50$\,GeV, $m_{H,H^\pm}=250$\,GeV, and $t_\beta=50$. }
		\label{fig:ms-k1}
	\end{center}
\end{figure}

In our scenario with $m_A<m_h<m_{H,H^\pm}$ and $t_\beta\gtrsim30$, one can read from 
the DM couplings (\ref{kappas}) that the main DM annihilation channels depending on $m_S$ can be categorized simply by $SS\to \tau\bar\tau$ for 
$m_S<m_A$; $SS\to AA$ for $m_S>m_A$, and $SS\to AA, HH/H^+ H^-$ for $m_S>m_{H,H^\pm}$.
For our analysis, we take a representative parameter set: $m_A=50$ GeV, $m_{H,H^\pm}=250$ GeV, and
$t_\beta=50$.
 

First, in case of $m_S<m_A$, the DM pair annihilation  goes through $SS \to h^*/H^* \to 
\tau \bar\tau$, leading to the corresponding annihilation rate:
\beq
\sigma v_{rel} (SS\to \tau \bar\tau) \!=\! \frac{ m_\tau^2 }{4\pi} 
\left|  \frac{\kappa_h}{P_h}+
\frac{\kappa_H t_\beta}{P_H}\right|^2 \!\!
\left( 1- \frac{m_\tau^2}{m_S^2}\right)^{3/2}\!\!\!\!\!\!,
\eeq
where $P_{h,H} \equiv 4 m_S^2 - m_{h,H}^2 + i \Gamma_{h,H}\, m_{h,H}$.
Away from the resonance point, 
the thermal freeze-out condition,  $\sigma v_{rel} \approx
 2 \times 10^{-9}\, \mbox{GeV}^{-2}$, is satisfied by 
\beq \label{eq:relic1}
\big| \kappa_h + \kappa_H t_\beta \frac{m_h^2}{m_H^2} \big| \approx 1.45\,.
\eeq
Considering the required limit of $\kappa_1 \gg \kappa_2$ (and thus $\kappa_H \approx \kappa_1/t_\beta$), Eq.~(\ref{eq:relic1}) requires
\beq
|\kappa_1 | \approx 5.8 \left( m_H \over 250\,\mbox{GeV}\right)^2 \,.
\eeq
This behavior is shown by the red curve for $m_S<50$ GeV in Fig~\ref{fig:ms-k1}, which is however disfavored by
the recent Fermi-LAT detection of gamma rays from dwarf galaxies \cite{Ackermann:2015zua}.

For $m_S>m_A$, the $SS\to AA$ channel is the dominant annihilation process leading to  
\begin{align}
&\sigma v_{rel}(SS\to AA) = \frac{1}{16\pi m_S^2} \sqrt{1-\frac{m_A^2}{m_S^2}} \nn \\ &\hspace*{1cm}\times  \bigg(\kappa_{AA}+ \frac{\kappa_h \lambda_{hAA}\, v^2 (4 m_S^2 - m_h^2)}{|P_h|^2} \nn \\
&\hspace*{1cm}+ \frac{\kappa_H \lambda_{HAA}\, v^2 (4 m_S^2 - m_H^2)}{|P_H|^2} \bigg)^2,
\end{align}
where in the alignment limit the triple scalar couplings are given by
\begin{align}
\lambda_{hAA} &= \frac{\left(m_h^2 - 2 m_A^2\right) \left(c_\beta^2 - s_\beta^2 \right)}{v^2}, \\
\lambda_{HAA} &= \frac{1}{v^2} \bigg[ m_H^2 s_\beta^2 \left(1 + t_\beta \right)  - m_{12}^2\left(\frac{1 }{c_\beta^2}+ \frac{1 }{s_\beta^2}\right) \nn \\
&+ 4 \, m_A^2 c_\beta s_\beta \bigg].
\end{align} 
%
%

The curve satisfying relic density with the mentioned annihilation mode can be seen  from Fig.~\ref{fig:ms-k1} for the range 50\,GeV\,$<m_S<$\,250\,GeV. As discussed in Sec.~\ref{sec:model} that in the large $t_\beta$ limit $\kappa_h \simeq \kappa_2$, the resonance behavior at $m_S=m_h/2$ is absent in $m_S-\kappa_1$ plane. It can also be seen that due to $\lambda_{HAA} >\lambda_{hAA}$, a huge enhancement of annihilation cross section near the $H$ resonance region rendering tiny values of $\kappa_1$ to obtain the observed relic density.  
 
For $m_S>m_{H,H^\pm}$,   the  $SS \to HH, H^+ H^-$ channels are open to give additional contribution given as
\begin{align}
&\sigma v_{rel}(SS\to HH/H^+ H^-) = \frac{3}{16\pi m_S^2} \sqrt{1-\frac{m_H^2}{m_S^2}} \nn \\ &\hspace*{1cm}\times  \Big(\kappa_{AA}  + \frac{\kappa_h \lambda_{hH^+H^-}\, v^2 (4 m_S^2 - m_h^2)}{|P_h|^2} \nn \\
&\hspace*{1cm}+ \frac{\kappa_H \lambda_{HH^+H^-}\, v^2 (4 m_S^2 - m_H^2)}{|P_H|^2} \Big)^2,
\end{align}
assuming $m_H=m_{H^\pm}$. The triple scalar couplings at the alignment limit are
\begin{align}
\lambda_{hH^+H^-} &=  \frac{\left(m_h^2 - 2 m_H^2\right) \left(c_\beta^2 - s_\beta^2 \right)}{v^2},\\
\lambda_{HH^+H^-} &= \frac{1}{v^2} \bigg[ m_H^2 s_\beta^2 \left(1 + t_\beta  + \frac{4}{c_\beta} \right)  \nn \\
&- m_{12}^2\left(\frac{1 }{c_\beta^2}+ \frac{1 }{s_\beta^2}\right)  \bigg].
\end{align}
The total effect of all three annihilation channels namely $SS\to \tau \tau, \,AA,\,HH/H^+H^-$ in the analysis is depicted in Fig.~\ref{fig:ms-k1} for the range $m_S>250$\,GeV where the observed relic density is easily obtainable with $\kappa_1 \simeq\mathcal{O}(10^{-1})$.

Fermi-LAT gamma ray detection from dwarf galaxies put strong bounds on the annihilation rates for the $2 \tau$ (Fig.~1 in Ref.~\cite{Ackermann:2015zua}) and $4\tau$ (Fig.~9 in Ref.~\cite{Elor:2015bho}) final states.  Both of them are similar, disfavoring $m_S\lesssim 80$ GeV. In Fig.~\ref{fig:ms-k1}, we show the excluded parameter space in gray shaded region. It should be noted that the indirect bound shown here is imposed in a conservative way assuming 100\% branching fraction for $H$ and $H^\pm$ to $\tau$ states and still leaves the region $m_S\ge 80\,$GeV completely accessible.
In principle, one has to consider the decay channels $H\to A Z$ and $H^\pm \to A W^\pm$ leading to one more step for the tau productions. However, it does not put a meaningful bound for $m_S>m_{H,H^\pm}$ as it slightly modifies the gamma ray bound
which can be found from Fig.~9 of Ref.~\cite{Elor:2015bho}.

\section{Conclusion}
\label{sec:conclusion}

In this work we consider an extension of the SM with an additional $SU(2)_L$ Higgs doublet and with a 
singlet scalar serving as a viable DM candidate. Our  particular interest is in the 2HDM of Type-X which 
can explain muon $g-2$ anomaly in the parameter space allowing a light pseudo-scalar $A$
and large $\tan\beta$,  and thus provides interesting testable signatures at the LHC. 
This scenario reveals a simple characteristic of the allowed parameter space consistent with 
the observed DM relic density and various constraints from direct and indirect detections.

The strong constraint on the SM Higgs portal scenario from direct detection experiments is evaded in a distinguishing way by extra Higgs portal present in the model. The recent XENON1T limit and 
the future sensitivity of XENONnT and LUX-ZEPLIN experiments severely constrains 
the quartic coupling $\kappa_2$ of the DM to one of the Higgs doublets (mostly SM-like) 
whereas the coupling $\kappa_1$ to other Higgs doublet is permitted up to $\mathcal{O}(1)$ values.

Such freedom allows us to obtain the correct relic density in the parameter space where muon $g-2$ anomaly can be explained. In this region of parameter space, the relevant annihilation channels for the DM pair are $\tau \tau, \,AA,\,HH/H^+H^-$.   As the DM annihilation leads to the $2\tau$ or $4\tau$ final state, Fermi-LAT data from gamma ray detection exclude the DM mass below about 80 GeV. We find that  the relic density can be obtained with reasonable values of the coupling $\kappa_1$
 for the DM mass opening up the annihilation channel of $AA$.


\section*{Acknowledgments }
EJC thanks the Galileo Galilei Institute for Theoretical Physics (GGI) for the hospitality and discussions within the program ``Collider Physics and the Cosmos''.

\end{document}